\documentclass[12pt]{article}
\usepackage[dvips]{color}
\usepackage{amsmath}
\usepackage{graphicx}
\usepackage{subfigure}
\usepackage{epsfig}
\usepackage{amsmath}
\usepackage{cite}
\usepackage{color}
\usepackage{subfigure}
\usepackage{multirow}

\begin{document}
\begin{center}
\Large{\bf Topological classification and black hole thermodynamics }\\
\small \vspace{1cm}{\bf Mohammad Reza Alipour $^{\star}$\footnote {Email:~~~mr.alipour@stu.umz.ac.ir}},  \quad
 {\bf Mohammad Ali S. Afshar $^{\star}$\footnote {Email:~~~m.a.s.afshar@gmail.com}},  \quad{\bf Saeed Noori Gashti$^{\dag,\star}$\footnote {Email:~~~saeed.noorigashti@stu.umz.ac.ir}},\quad
{\bf Jafar Sadeghi $^{\star}$\footnote {Email:~~~pouriya@ipm.ir}}\\
\vspace{0.5cm}$^{\star}${Department of Physics, Faculty of Basic
Sciences,\\
University of Mazandaran
P. O. Box 47416-95447, Babolsar, Iran}\\
\vspace{0.2cm}$^{\dag}${School of Physics, Damghan University, P. O. Box 3671641167, Damghan, Iran}

\end{center}
\begin{abstract}
One of the new methods that can be used to study the thermodynamics, critical points of a system based on a topological approach is the study of topological charges using Duan's $\phi$-mapping method. In this article, we will attempt to use this method to study three different black holes, each with different coefficients in their metric function, in order to determine the class of critical points, these black holes have in terms of phase transition. Through this analysis, we found that the Euler-Heisenberg black hole has two different topological classes, and the parameter $"a"$ added to the metric function by QED plays an important role in this classification.
While a black hole with a non-linear electrodynamic field, despite having an electromagnetic parameter, which is added to its metric function, has only one topological class, and its $"\alpha"$ parameter has no effect on the number of critical points and topological class.
Finally, the Yang-Mills black hole in massive gravity will have a different number of critical points, depending on the coefficient $"c_i"$, which is related to massive gravity and leads to different topological classes. However, this black hole exhibits the same phase structure in all cases.\\\\
Keywords: Critical Point, Topological Charge, Thermodynamics of Black holes
\end{abstract}
\newpage
\tableofcontents

\section{Introduction}
A new method for studying the thermodynamics and phase transitions of black holes in Anti-de Sitter space has been created. The method demonstrates a precise connection between the phase transition of charged AdS black holes from small to large size and the phase transition of ordinary fluids from liquid to gas \cite{1}.
This also involves a critical area where the phase transition shifts from being first-order to second-order \cite{1,2,3}. This critical region is particularly fascinating since it displays certain universal characteristics that are dependent upon the scaling of thermodynamic quantities.
A recent proposal has suggested utilizing a topological approach, based on Duan's $\phi$-mapping theory, to determine the characteristics of critical points for certain types of black holes that possess electric charges and nonlinear electromagnetic fields \cite{4,5,6}.

Wei et al. presented two different approaches to study topological thermodynamics based on temperature and generalized free energy function. The first method involves studying the temperature function, and then constructing the potential based on these assumptions. In the second approach, with the assumption that black holes can be considered as defects in the thermodynamic parameter space, their solutions can be investigated using the generalized off-shell free energy. In that case, the stability and instability of the obtained black hole solutions are determined by positive and negative winding numbers, respectively \cite{aa,bb,cc,dd,ee,ff,gg,hh,ww}.

It is important to determine if this approach can be applied to more complex scenarios with multiple critical points, not just limited to the liquid-gas phase transition \cite{7}. Apart from the transition of black holes from small to large, there are other fascinating phenomena that are similar to conventional thermodynamic systems. These include phase transitions that occur multiple times in the same direction as a parameter is changed, or those that involve more than two phases, such as solid-liquid-gas transitions.
These phenomena occur in various types of black holes, particularly in higher dimensions, and involve additional terms in the gravitational equations that rely on higher powers or derivatives of the curvature \cite{8,9,10,11}. Thermodynamics in extended phase space pertains to a framework utilized for analyzing the thermodynamics and phase transitions of black holes in Anti-de Sitter (AdS) space. In this framework, the cosmological constant is regarded as a dynamic variable and interpreted as thermal pressure.
In this framework, the mass parameter of the black hole is viewed as enthalpy instead of internal energy. This allows for more intricate and diverse phenomena, such as Van der Waals-like phase transitions, reentrant phase transitions, superfluidity, polymer-like phase transitions, triple points, and new dual relationships. The formula for pressure in this extended phase space is as follows: $P = -\Lambda/8\pi$, where $\Lambda$ represents the cosmological constant \cite{12,13,14,15}.
An application of this framework involves examining the holographic dualities that exist between black holes in AdS space and conformal field theories on the boundary. For instance, the Hawking-Page phase transition between Schwarzschild AdS black holes and thermal AdS space is linked to the confinement/deconfinement transition in the boundary gauge theory \cite{16,17,18}. Another use is the creation of holographic heat engines via the utilization of black holes in AdS space as working substances.
These processes are cyclical and can convert heat into work or vice versa, making them suitable for assessing the efficiency and effectiveness of different thermodynamic cycles. A third application involves examining the microstructure of black holes in AdS space using thermodynamic geometry techniques \cite{19,20}. This approach enables the investigation of fluctuations and correlations of thermodynamic variables near critical points and can unveil universal characteristics of black hole phase transitions.\\
Wei and Liu state that it is possible to use a combination of classical thermodynamic functions and Duan's $\phi$-mapping method to achieve a new approach to studying the thermodynamic structures of black holes \cite{5}. This method can also easily determine the possibility of a phase transition at the critical points of the system being studied. Based on this, it is claimed that critical points with a topological charge of $+1$ $(Q_t=+1)$ do not exhibit a first-order phase transition, which is considered "novel". On the other hand, points with a topological charge of $-1$ $(Q_t=-1)$ can be considered to have a first-order phase transition, which is referred to as "conventional". But in \cite{5}, this proposal was doubted by examining the study of charged Gauss-Bonnet black holes in AdS spacetime. They showed that there are conventional critical points that do not cause first-order phase transition, so condition Wei cannot be considered a necessary and sufficient condition for classifying points for the phase transition structure. Therefore, to solve this problem, the critical points were reclassified in another way. In this way, as pressure increases, the novel critical point will be the point that results in a new phase (whether stable or unstable). In contrast, the conventional point will be equivalent to the point where the change of phases disappears \cite{21,22}. It is mentioned in \cite{5} that changes in topological charge class can be caused by either gravity or gauge corrections. Of course, it is suggested in the article \cite{21} that gravitational corrections have no effect on changing the topological class of a black hole (Gauss-Bonnet black hole), while gauge corrections can change the topological class (Born-Infeld black hole) \cite{22}. However, when both factors are present, they will not affect the topological charge class. On the other hand, the application of the "Wei" method to the Lovelock black hole has also yielded other results, which demonstrate that for a charged Lovelock black hole with arbitrary dimensions, a certain class of topological charge can exhibit a different phase structure \cite{23}.
Furthermore, the uncharged Lovelock black hole in dimensions $d=7$ and $d\geq8$ can have different phase structures depending on the class of topological charge.
Therefore, it can be concluded that a change in the topological charge class may indicate the presence of changes in the phase structure, but the reverse is not necessarily true \cite{23}.
According to the aforementioned experiences, we selected black holes that had gravitational and gauge corrections, as well as different dimensions. We attempted to investigate the claims and discovered some intriguing points.
The Euler-Heisenberg black hole, which has electromagnetic gauge corrections due to the coefficient $"a"$ in its metric function, confirms the previous results regarding the effect of gauge corrections on the topological charge class in the study of the Born-Infeld black hole \cite{22}.
However, a black hole with a non-linear electrodynamics field does not exhibit any effects despite having gauge corrections that appear with $"\alpha"$ coefficient in its metric function.
Another interesting point is that the Yang-Mills black hole in massive gravity, which has both gravitational and gauge corrections, behaves differently because it is the gravitational corrections that change the topological class.
Also, the Yang-Mills black hole in massive gravity challenges the proposal presented in \cite{23} because our results show that despite belonging to a different topological charge class, it displays the same phase structure.
Finally, we can state that although the "Wei" method can be a simple and quick way to make an initial comment about the existence of a first-order phase transition at critical points.
But this method still cannot be considered a reliable way to make a final statement, and the conventional method of thermodynamics should still be used to verify the first-order phase transition at the critical point.
Therefore, the above content motivated us to organize this article in the following form.\\
In section 2, we reviewed the thermodynamics of the considered black holes.
In section 3, we investigated the relationship between the critical points of the black hole and its topological charge.
In section 4, we calculated the critical points and topological charges related to the considered black holes and stated the related phase structure.
Finally, we expressed our conclusion in section 5.
\section{ Thermodynamics of black holes}
Black hole thermodynamics is the study area that aims to reconcile the laws of thermodynamics with the existence of black hole event horizons. Researchers consider various approaches, including the study of black holes and their thermodynamic characteristics, as a promising tool to gain insight into quantum gravity. A considerable amount of research has been conducted on the thermodynamics of black holes, and for further review, refer to \cite{a,b,c,d,e,f}.
We first review thermodynamic quantities of the black hole in the extended space (considering the cosmological constant $\Lambda$ as the pressure $P$ and the thermodynamic volume $V$ as its conjugate quantity). In the following, we will examine the thermodynamic quantities of black holes in Euler-Heisenberg, nonlinear electrodynamics, and Yang-Mills' five-dimensional massive gravity, respectively. Also, we set $G=\hbar=c=1$.
\subsection{Case I: Euler-Heisenberg-AdS black hole}
Euler Heisenberg's black hole, which has spherical symmetry, its metric function is obtained \cite{24},
\begin{equation}\label{1}
\begin{split}
f(r)=1-\frac{2M}{r}+\frac{Q^2}{r^2}-\frac{\Lambda r^2}{3}-\frac{a Q^4}{20 r^6},
\end{split}
\end{equation}
where $M$ and $Q$ are the mass and electric charge of the black hole, respectively. Additionally, $\Lambda$ is a cosmological constant whose expanded phase space is interpreted as pressure $P=-\frac{\Lambda}{8\pi}$, and $a$ is related to the QED parameter. When $a=0$, it is reduced to charged AdS black hole. The thermodynamic quantities for this black hole can be obtained from \cite{24},
\begin{equation}\label{2}
\begin{split}
T=\frac{1}{4\pi r_h}-\frac{Q^2}{4\pi r_h^3}+\frac{aQ^4}{16\pi r_h^7}+2Pr_h,
\end{split}
\end{equation}
\begin{equation}\label{3}
\begin{split}
S=\pi r_h^2,
\end{split}
\end{equation}
\begin{equation}\label{4}
\begin{split}
\Phi_{em}=\frac{Q}{r_h}\left(1-\frac{aQ^2}{10r_h^4}\right),
\end{split}
\end{equation}
\begin{equation}\label{5}
\begin{split}
M=\frac{r_h}{2}\left(1+\frac{Q^2}{r_h^2}+\frac{8\pi P}{3}r_h^2-\frac{aQ^4}{20r_h^6}\right).
\end{split}
\end{equation}
In the above relations, $T$, $S$, $\Phi_{em}$, and $r_h$ are the Hawking temperature, entropy, electromagnetic potential, and the radius of the black hole event horizon, respectively.
\subsection{Case II: Black hole with non-linear electrodynamic field}
When we add a non-linear electrodynamic field correction term to Einstein-Maxwell's theory, the black hole metric is obtained as follows \cite{25},
\begin{equation}\label{6}
\begin{split}
f(r)=1+2\sqrt{\alpha}Q-\frac{2M}{r}+\frac{Q^2}{r^2}+\frac{ r^2}{3}\left(\frac{8\pi}{3}P-\alpha\right),
\end{split}
\end{equation}
where $\alpha$ is the coupling constant. When $\alpha=0$, it reduces to Einstein-Maxwell theory. Furthermore, we can find the thermodynamic properties of this black hole in \cite{25},
\begin{equation}\label{7}
\begin{split}
T=\frac{1}{4\pi r_h}(1+2\alpha Q)-\frac{Q^2}{4\pi r_h^3}+\frac{r_h}{4\pi}(8\pi P-\alpha),
\end{split}
\end{equation}
\begin{equation}\label{8}
\begin{split}
S=\pi r_h^2,
\end{split}
\end{equation}
\begin{equation}\label{9}
\begin{split}
\Phi_{em}=\frac{Q}{r_h}\left(1+\sqrt{\alpha}r_h^2\right),
\end{split}
\end{equation}
\begin{equation}\label{10}
\begin{split}
M=\frac{r_h}{2}\left(1+\frac{Q^2}{r_h^2}+\frac{8\pi P}{3}r_h^2-\frac{\alpha }{3}r_h^2+2\sqrt{\alpha}Q\right).
\end{split}
\end{equation}
We can also obtain the Gibbs free energy ($G$) by using the relations above and $G=M-TS$.
\subsection{Case III: Five-dimensional Yang–Mills black holes in massive gravity}
In this section, we introduce the five-dimensional Yang-Mills black hole in massive gravity. Its metric is given by \cite{26},
\begin{equation}\label{11}
\begin{split}
&f(r)=\frac{4\pi P r^2}{3}\left[1-\left(\frac{r_h}{r}\right)^4\right]-\frac{2e^2}{r^2}\ln\frac{r}{r_h}\\
&+\frac{m^2 c_0 c_1 r}{3}\left[1-\left(\frac{r_h}{r}\right)^3\right]+m^2 c_0^2 c_2 \left[1-\left(\frac{r_h}{r}\right)^2\right]+\frac{2m^2c_0^3c_3}{r}\left[1-\left(\frac{r_h}{r}\right)\right],
\end{split}
\end{equation}
where $e$, $m$, and $c_{\nu}$ $(\nu=0,1,2,3)$ are coupling constants of the Yang-Mills theory, the mass, and constants term of massive gravity, respectively. Also, its thermodynamic quantities are as follows \cite{26},
\begin{equation}\label{12}
\begin{split}
T=\frac{1}{2\pi r_h^3}\left(\frac{8\pi P r_h^4}{3}-e^2\right)+\frac{m^2 c_0 }{2\pi r_h^2}\left(\frac{c_1 r_h^2}{2}+c_0c_2r_h+c_0^2c_3\right),
\end{split}
\end{equation}
\begin{equation}\label{13}
\begin{split}
M=\frac{3}{16\pi }\left(\frac{4\pi P r_h^4}{3}-2e^2\ln\frac{r_h}{L}\right)+\frac{3m^2c_0r_h}{16\pi}\left(\frac{c_1 r_h^2}{3}+c_0c_2r_h+2c_0^2c_3\right),
\end{split}
\end{equation}
\begin{equation}\label{14}
\begin{split}
S=\frac{r_h^3}{4},
\end{split}
\end{equation}
where $L$ is an arbitrary length constant that is considered to have a dimensionless argument of a logarithmic term.

\section{Topology of black hole thermodynamics}
In the extended thermodynamic framework, in order to study the critical points of a thermodynamic system such as a black hole, the temperature is rewritten as a function of pressure, entropy, and $x_i$, i.e., $T=T(P,S,x_i)$. Then, we use the following relationship to obtain the critical points \cite{5,21,22,23},
\begin{equation}\label{15}
\begin{split}
\left(\frac{\partial T}{\partial S}\right)_{P,x_i}=0, \qquad  \left(\frac{\partial^2 T}{\partial S^2}\right)_{P,x_i}=0.
\end{split}
\end{equation}
Recently, it has been shown in \cite{5} that each critical point can have a topological charge, which can be divided into two types: conventional and novel. In the following, we can eliminate the parameter $P$ using the relation $\left(\frac{\partial T}{\partial S}\right)_{P,x_i}=0$. Additionally, to check the topological charge of critical points, we define the thermodynamic function as follows \cite{5,21,22},
\begin{equation}\label{16}
\begin{split}
\Phi=\frac{1}{\sin\theta}T(S,x_i),
\end{split}
\end{equation}
where $\frac{1}{\sin\theta}$ is considered an auxiliary factor to simplify the analysis of topological calculations, we can then define a new vector field $\phi=(\phi^S,\phi^\theta)$ using Doan's $\phi$-mapping theory \cite{5,21,22} and we have,
\begin{equation}\label{17}
\begin{split}
\phi^S=\left(\frac{\partial \Phi}{\partial S}\right)_{\theta,x_i}, \qquad \phi^\theta= \left(\frac{\partial \Phi}{\partial \theta}\right)_{S,x_i}.
\end{split}
\end{equation}
The vector field $\phi$ is always zero at the point $\theta=\frac{\pi}{2}$, allowing for easy identification of the critical points in the thermodynamic system. Additionally, the points $\theta=0$ and $\theta=\pi$, where the field vector is perpendicular to the horizontal lines, act as boundaries in the parameter space.
The proposed structure has an important feature: when the vector field $\phi^a$ becomes zero (i.e., $\phi^a(x^i)=0$), its topological current $j^{\mu}$ becomes non-zero. As a result, we can calculate its topological charge in the parameter region $\Sigma$ as follows \cite{27,28},
\begin{equation}\label{18}
\begin{split}
Q_t=\int_{\Sigma}j^0 d^2x=\Sigma_{i=1}^N \omega_i,
\end{split}
\end{equation}
where $\omega_i$ is the winding number for $\phi^a(x^i)=0$. Since the winding number can be either $+1$ or $-1$, $Q_t$ can be positive or negative at critical points. This is suggested in \cite{5}, which divides them into two categories: conventional charge when $Q_t=-1$ and novel charge when $Q_t=+1$.
Also, to obtain the topological charge, we can use the contour $C$ around the critical points in the $\sqrt{S}-\theta$ or $r-\theta$ plane where the $n$ vector field is drawn, and obtain the normalization of the vector field,
\begin{equation}\label{19}
\begin{split}
n=(\frac{\phi^S}{\|\phi \|}, \frac{\phi^\theta}{\|\phi \|}).
\end{split}
\end{equation}
On the other hand, contour $C$ should be parametrized by $\vartheta\in (0,2\pi)$,
\begin{equation}\label{20}
\begin{split}
r=a \cos\vartheta+r_0,  \qquad \theta=b\sin\vartheta+\frac{\pi}{2},
\end{split}
\end{equation}
which $(r_0,\pi/2)$ is the center of the counter. Now, according to \eqref{19} and \eqref{20} , we define a new quantity that measures the deflection of the vector field along the contour. So, we will have,
\begin{equation}\label{21}
\begin{split}
\Omega(\vartheta)=\int_{0}^{\vartheta}\epsilon_{ij}n^i\partial_{\vartheta}n^{j}d\vartheta,
\end{split}
\end{equation}
where $i,j= S, \theta$. Then the topological charge must be $Q_t=\frac{\Omega(2\pi)}{2\pi}$.

\section{The models}
First, we attempt to find the critical points of black holes using equations \eqref{15}, \eqref{16}, and \eqref{17}. Next, we determine the corresponding topological charge for these critical points using equations \eqref{18}, \eqref{19}, \eqref{19}, and \eqref{20}. Finally, we analyze their properties by creating diagrams.
\subsection{Case I}
\begin{figure}[h!]
 \begin{center}
 \subfigure[]{
 \includegraphics[height=5cm,width=5cm]{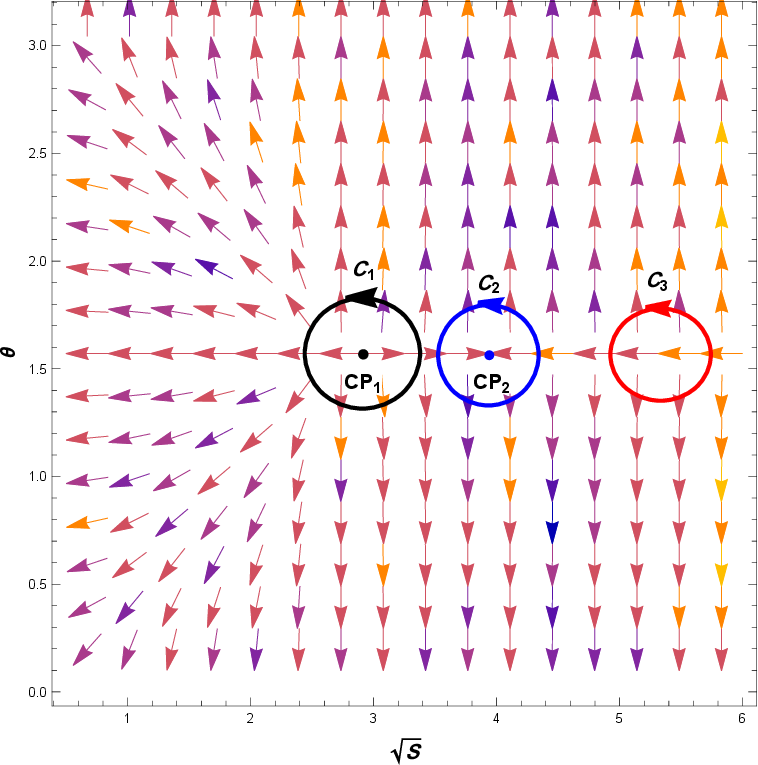}
 \label{fig1a}}
 \subfigure[]{
 \includegraphics[height=5cm,width=5cm]{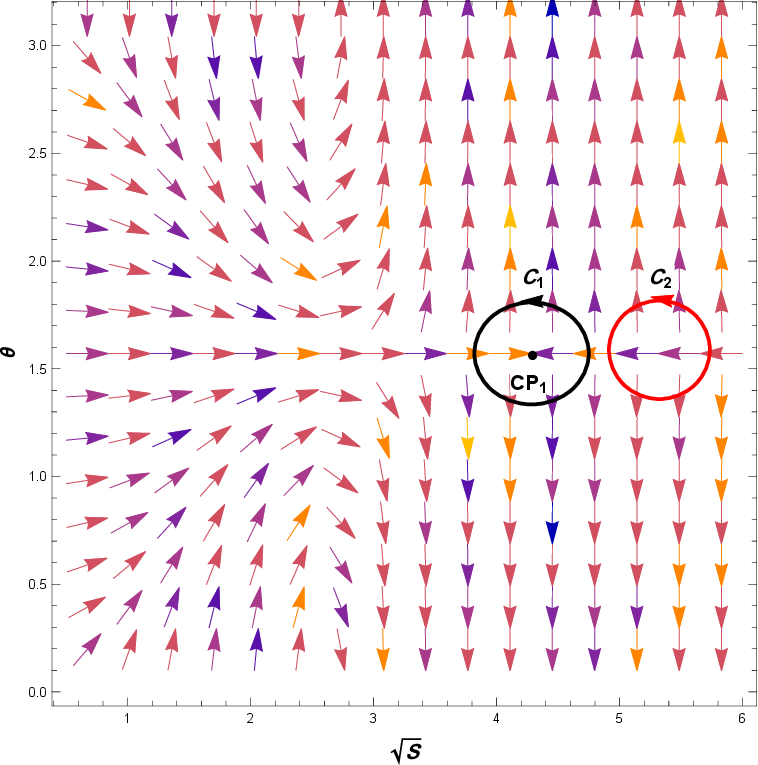}
 \label{fig1b}}
 \subfigure[]{
 \includegraphics[height=5cm,width=5cm]{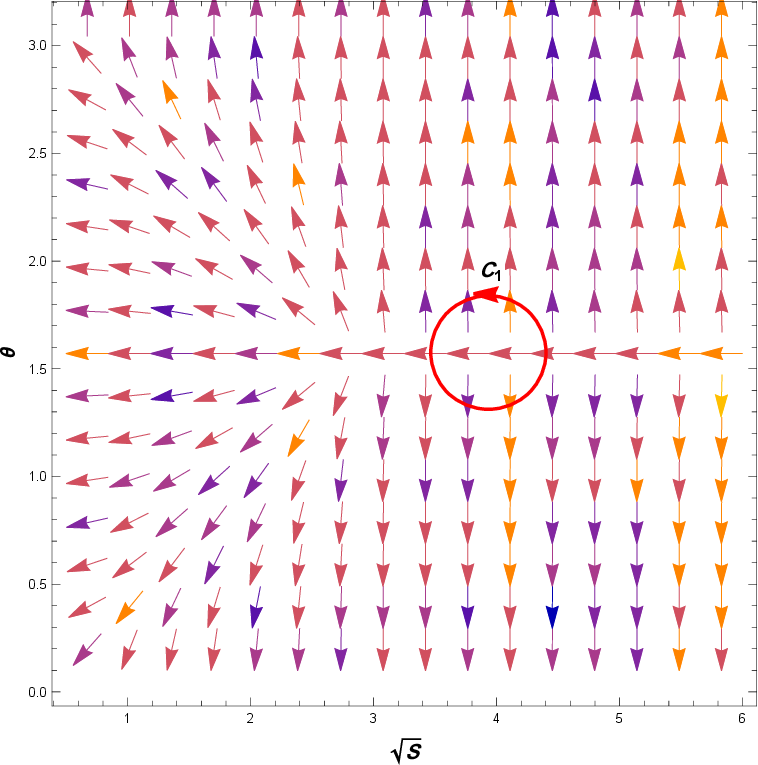}
 \label{fig1c}}
 \caption{\small{The  arrows represent the vector field n on  the $\sqrt{S}-\theta$ plane for the Euler-Heisenberg black hole with the charge $Q=1$ and different values for parameter $a$. So that, (a) $a= 3$, (b) $a= -1$, (c) $a= 5$.}}
 \label{fig1}
 \end{center}
 \end{figure}

\begin{figure}[h!]
 \begin{center}
 \subfigure[]{
 \includegraphics[height=5cm,width=5cm]{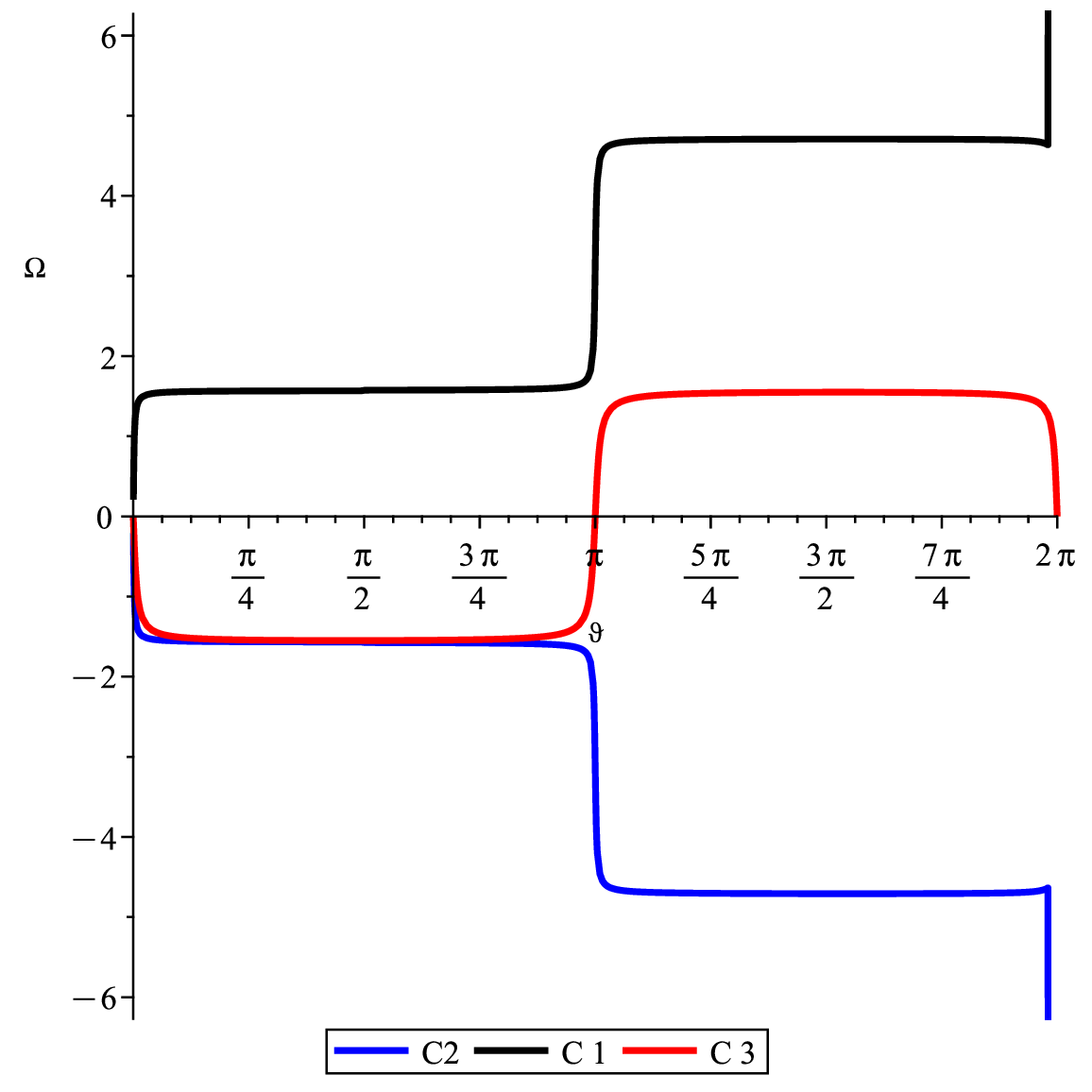}
 \label{fig2a}}
 \subfigure[]{
 \includegraphics[height=5cm,width=5cm]{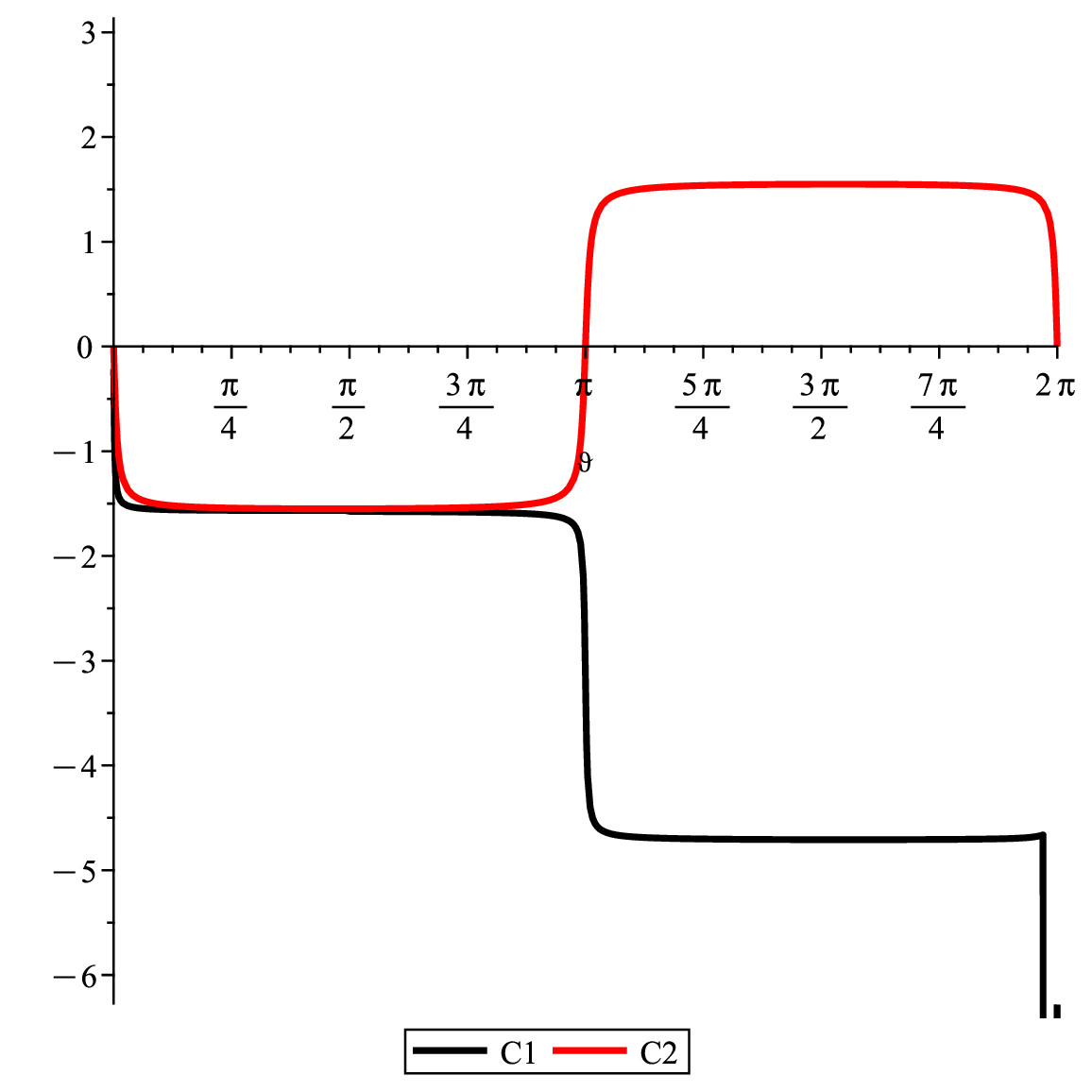}
 \label{fig2b}}
  \subfigure[]{
 \includegraphics[height=5cm,width=5cm]{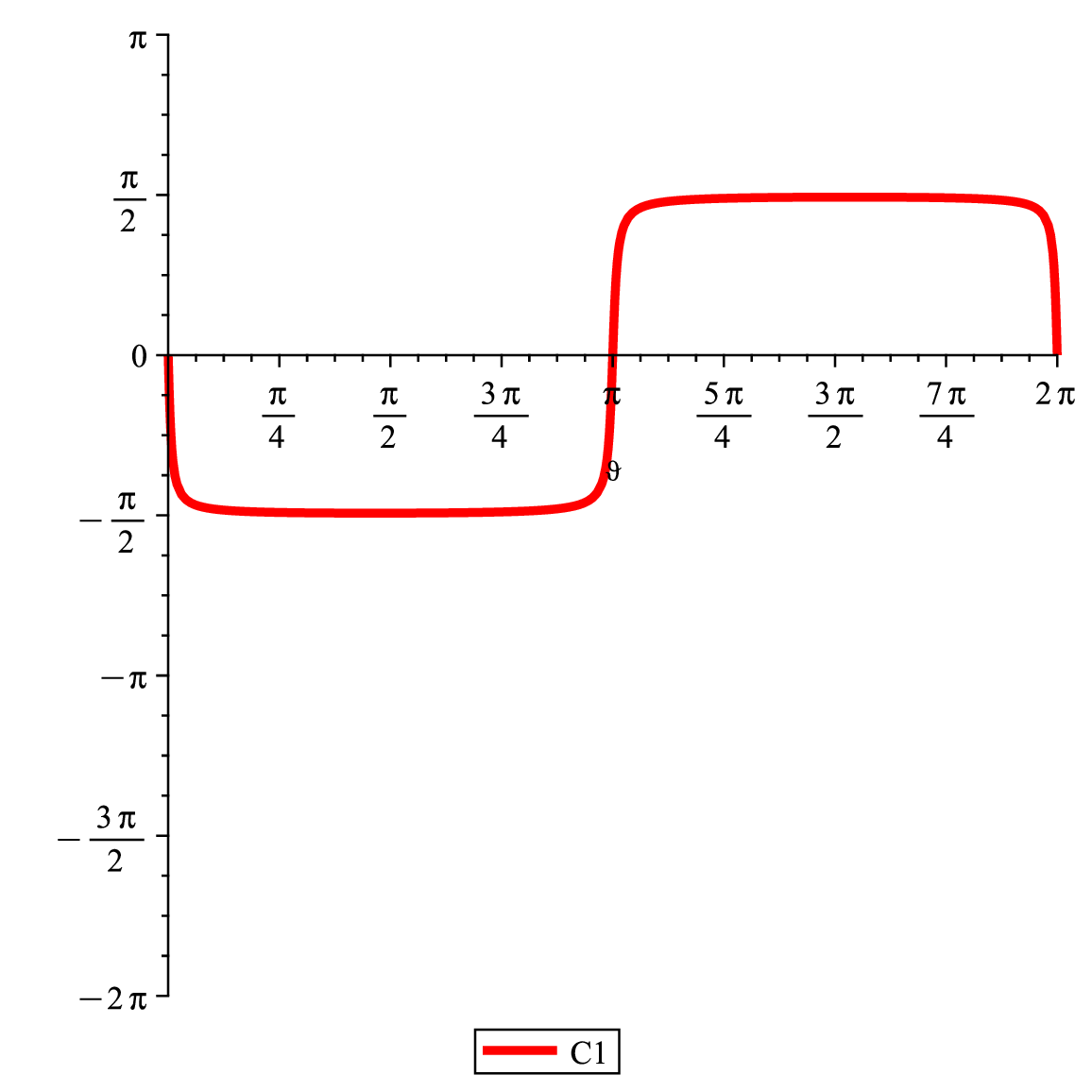}
 \label{fig2c}}
 \caption{\small{ $\Omega$ vs $\vartheta$ for contours $C$.}}
 \label{fig2}
 \end{center}
 \end{figure}

At first, we use relations \eqref{2} and \eqref{15} to eliminate the parameter $P$, resulting in,
\begin{equation}\label{22}
\begin{split}
T =\frac{a \,Q^{4} \pi^{3}-2 \pi  Q^{2} S^{2}+S^{3}}{2 \sqrt{\pi}\, S^{\frac{7}{2}}}.
\end{split}
\end{equation}
Then, by substituting the equation \eqref{22} into \eqref{16}, the thermodynamic function $\Phi$ for the Euler-Heisenberg black hole can be determined as follows,
\begin{equation}\label{23}
\begin{split}
\Phi =\frac{a \,Q^{4} \pi^{3}-2 \pi  Q^{2} S^{2}+S^{3}}{2 \sin \! \left(\theta \right) S^{\frac{7}{2}} \sqrt{\pi}}.
\end{split}
\end{equation}
Now, with respect to equations \eqref{17}, and \eqref{23} we investigate the vector fields as,
\begin{equation}\label{24}
\begin{split}
&\phi^{S}=-\frac{7 a Q^{4} \pi^{3}-6 \pi  Q^{2} S^{2}+S^{3}}{4 \sin \left(\theta \right) S^{\frac{9}{2}} \sqrt{\pi}}\\
&\phi^{\theta}=-\frac{\left(a Q^{4} \pi^{3}-2 \pi  Q^{2} S^{2}+S^{3}\right) \cos \left(\theta \right)}{2 \sin \left(\theta \right)^{2} S^{\frac{7}{2}} \sqrt{\pi}}.
\end{split}
\end{equation}
According to relation \eqref{24}, we can obtain the critical points. Of course, the parameter "a" plays an important role in determining the number of critical points shown in Figure \eqref{fig1}.
 Also, the behavior of the deviation angle $\Omega(\vartheta)$, for the given contours in Figure \eqref{fig1}, is shown in Figure \eqref{fig2}.
As shown in Figure \eqref{fig1}, the parameter $"a"$ affects the number of critical points, and can be divided into the following three categories:
(1) When $0\leq a\leq\frac{32Q^2}{7}$, we have two critical points as shown in Figure \eqref{fig1a}. The topological charge corresponding to each critical point is $Q_{CP_2}= -1$ and $Q_{CP_1}= +1$. In this case, the total topological charge is $Q_{total}=Q_{CP_1}+Q_{CP_2}=0$.
(2) When $a < 0$, we have a critical point as shown in Figure \eqref{fig1b}, and the topological charge corresponding to it is $-1$, i.e. $Q_{CP_1}= -1$. So, the total topological charge is $Q_{total}=Q_{CP_1}=-1$.
(3) When $a>\frac{32Q^2}{7}$, we do not have a critical point and the topological charge for it.

\begin{figure}[h!]
 \begin{center}
 \subfigure[]{
 \includegraphics[height=5cm,width=5cm]{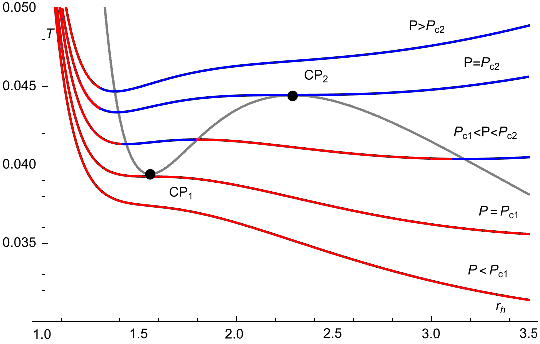}
 \label{3a}}
 \subfigure[]{
 \includegraphics[height=5cm,width=5cm]{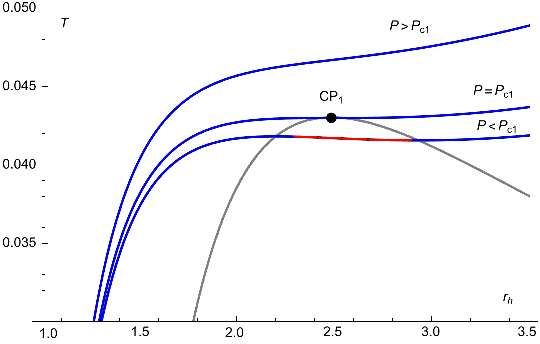}
 \label{3b}}
 \caption{\small{Isobaric curves (blue=stable, red=unstable) and the spinodal curve (gray line)
for Euler-Heisenberg black hole.  We have set $Q = 1$ and  (a) $a=3$, (b) $a= -1$.}}
 \label{fig3}
 \end{center}
 \end{figure}

\begin{figure}[h!]
 \begin{center}
 \subfigure[]{
 \includegraphics[height=5cm,width=5cm]{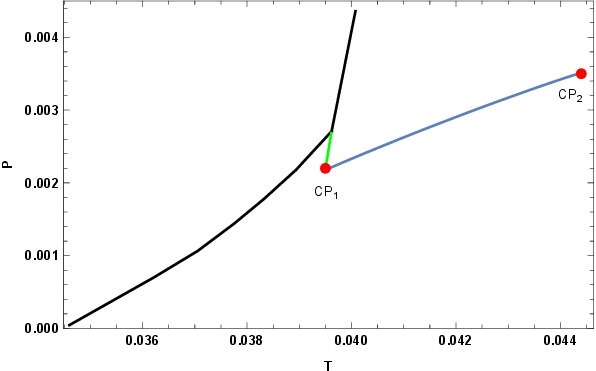}
 \label{fig4a}}
 \subfigure[]{
 \includegraphics[height=5cm,width=5cm]{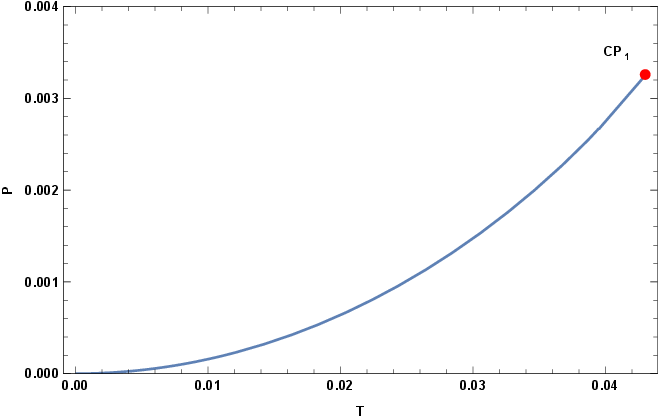}
 \label{fig4b}}
 \caption{\small{Phase diagram showing  for the  Euler-Heisenberg black hole. (a) The reentrant phase structure for the $a=3$ and $Q=1$. (b) The first-order phase transitions near the conventional critical point $CP_1$ for the $a=-1$ and $Q=1$.}}
 \label{fig4}
 \end{center}
 \end{figure}

We can see in Figure \eqref{fig3} that the novel critical point is at the minimum point of a part of the spinodal curve, while the conventional critical point is the maximum point of the spinodal curve.
As shown in Figure \eqref{fig3}, at the novel critical point, the black hole has an unstable state, but when the pressure is $P>P_{c1}$, a new phase of stability and instability appears in the black hole.
Also, the black hole is in a stable state at the conventional critical point, and the stability of the black hole is maintained by increasing the pressure. In this case, we will not have a new phase state. This result is in agreement with proposal in \cite{21}, where the novel critical point of the novel phase of stability and instability appears and disappears at the conventional critical point of the phase.
 We have found that the Euler-Heisenberg black hole falls into two distinct topological classes, resulting in two different phase structures similar to those of uncharged Lovelock AdS black holes with $d>7$ \cite{23}. As depicted in Figure \eqref{fig4a}, there is a reentrant phase transition comprising a first-order phase transition (blue curve) and a zeroth-order phase transition (green curve). However, in Figure \eqref{fig4b}, only the first-order phase transition is present.
\subsection{Case II}
\begin{figure}[h!]
 \begin{center}
 \subfigure[]{
 \includegraphics[height=5cm,width=5cm]{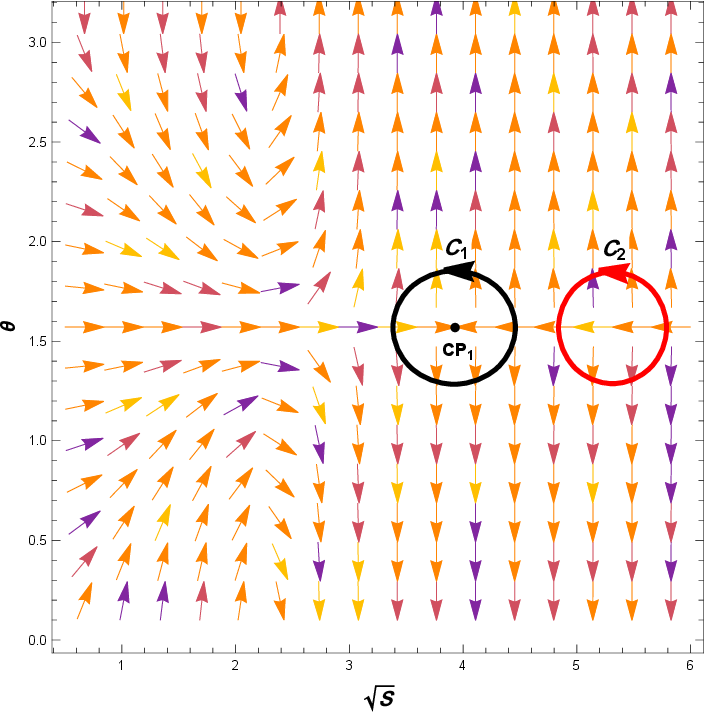}
 \label{fig5a}}
 \subfigure[]{
 \includegraphics[height=5cm,width=5cm]{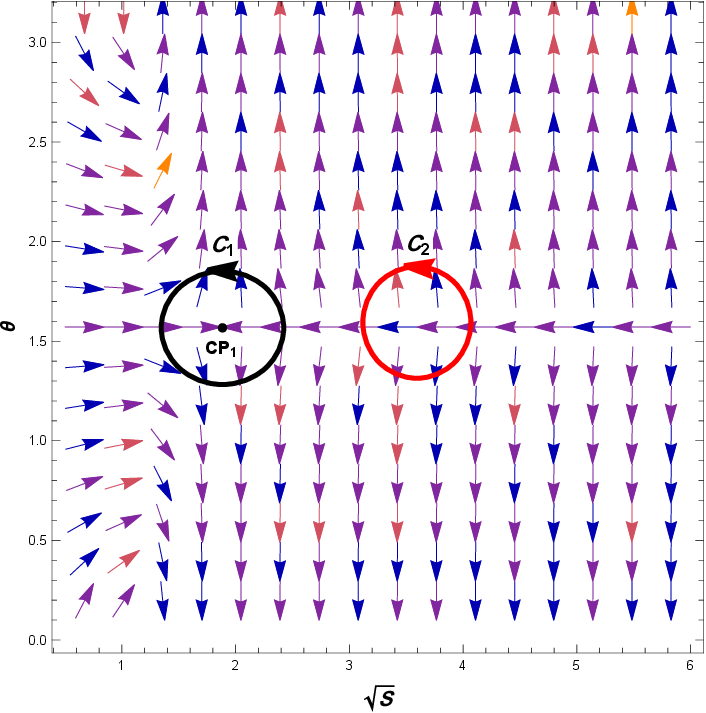}
 \label{fig5b}}
 \caption{\small{The arrows represent the vector field n on  the $\sqrt{S}-\theta$ plane for the nonlinear electrodynamics black hole with the charge $q=1$ and different values for parameter $\alpha$. So that, (a) $\alpha= 0.1$, (b) $\alpha= 2$.}}
 \label{fig5}
 \end{center}
 \end{figure}

\begin{figure}[h!]
 \begin{center}
 \includegraphics[height=5cm,width=5cm]{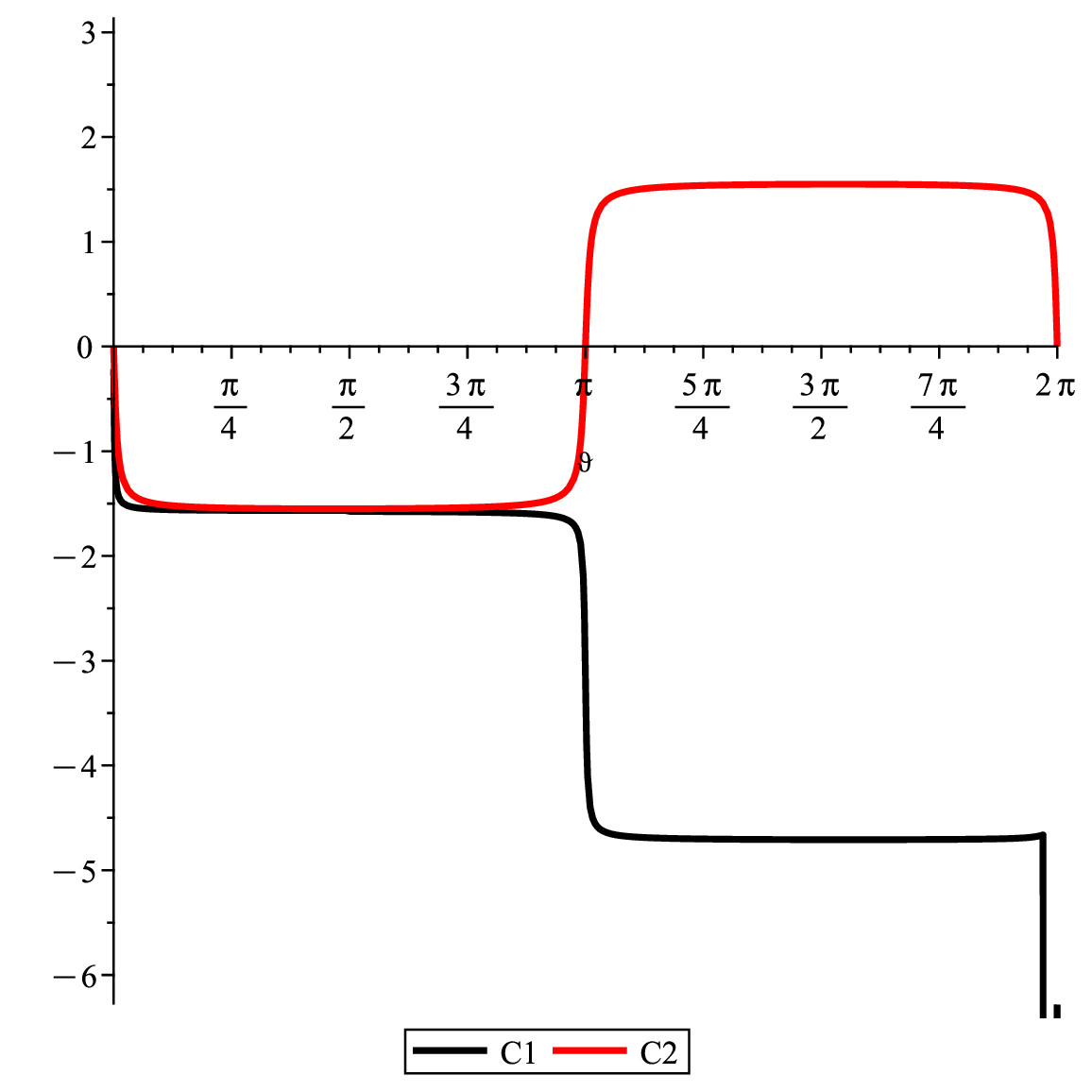}
 \caption{\small{ $\Omega$ vs $\vartheta$ for contours $C$.}}
 \label{fig6}
 \end{center}
 \end{figure}

Regarding equations \eqref{7} and \eqref{15}, we will have,
\begin{equation}\label{25}
\begin{split}
T =\frac{-2 q^{2} \pi +2 \alpha  q S +S}{2 S^{\frac{3}{2}} \sqrt{\pi}}.
\end{split}
\end{equation}
Also, by placing \eqref{25} in \eqref{16}, we get the thermodynamic function,
\begin{equation}\label{26}
\begin{split}
\Phi =\frac{-2 q^{2} \pi +2 \alpha  q S +S}{2 \sin \! \left(\theta \right) S^{\frac{3}{2}} \sqrt{\pi}}.
\end{split}
\end{equation}
In the following, we have the vector field components,
\begin{equation}\label{27}
\begin{split}
&\phi^{S}=-\frac{\alpha  q S -3 q^{2} \pi +\frac{1}{2} S}{2 S^{\frac{5}{2}} \sqrt{\pi}\, \sin \! \left(\theta \right)}\\
&\phi^{\theta}=-\frac{\left(-2 q^{2} \pi +2 \alpha  q S +S \right) \cos \! \left(\theta \right)}{2 \sin \! \left(\theta \right)^{2} S^{\frac{3}{2}} \sqrt{\pi}}.
\end{split}
\end{equation}
Now, we can obtain the critical points using equation \eqref{27}, which are illustrated in Figure \eqref{fig5} for various values of the parameter $"\alpha"$.
Also, the behavior of the deviation angle $\Omega(\vartheta)$, for the given contours in Figure \eqref{fig5}, is shown in Figure \eqref{fig6}.
Since the critical point at $(\sqrt{S}, \theta)=(\sqrt{\frac{6\pi q^2}{1+2\alpha q}}, \pi/2)$, we observe that critical points occur at smaller entropy or event horizons with increasing $\alpha$, as shown in Figure \eqref{fig5}. However, this does not alter the topological charge. Specifically, we have a conventional critical point for all values of $\alpha$, denoted by $e$, i.e. $Q_{total}=Q_{CP_1}=-1$, and this result is consistent with the charged AdS black hole in \cite{5}.

\begin{figure}[h!]
 \begin{center}
 \subfigure[]{
 \includegraphics[height=5cm,width=5cm]{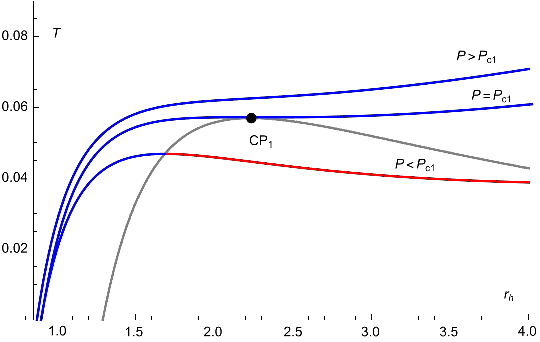}
 \label{1a}}
 \subfigure[]{
 \includegraphics[height=5cm,width=5cm]{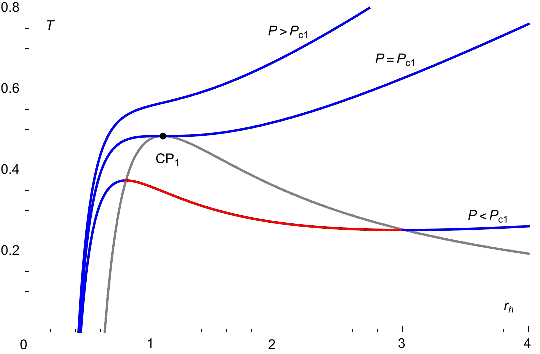}
 \label{1b}}
 \caption{\small{Isobaric curves (blue=stable, red=unstable) and spinodal curve (gray line)
for nonlinear electrodynamic black hole. We have set $q = 1$ and  (a) $\alpha= 0.1$, (b) $\alpha= 2$.}}
 \label{fig7}
 \end{center}
 \end{figure}

\begin{figure}[h!]
 \begin{center}
 \subfigure[]{
 \includegraphics[height=5cm,width=5cm]{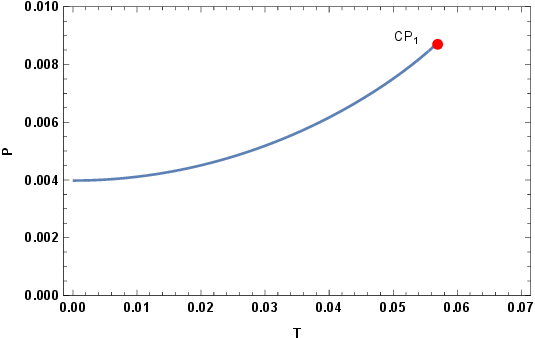}
 \label{fig8a}}
 \subfigure[]{
 \includegraphics[height=5cm,width=5cm]{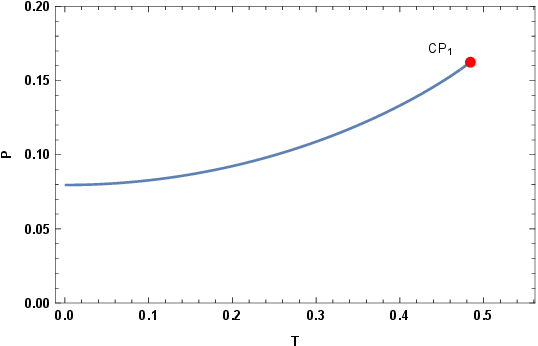}
 \label{fig8b}}
 \caption{\small{Phase diagram showing the nonlinear electrodynamic black hole. The first-order phase transitions occur near the conventional critical point ($CP_1$) for $q=1$ and (a) $\alpha=0.1$, (b) $\alpha=2$.}}
 \label{fig8}
 \end{center}
 \end{figure}

The nonlinear electrodynamics black hole is in the same topological class as charged Gauss-Bonnet AdS black holes \cite{21}, charged Lovelock AdS black holes \cite{23}, and charged AdS black holes \cite{5}. As shown in Figure \eqref{fig7}, when  $P<P_{c1}$, the black hole can exist in both stable and unstable states. With increasing pressure, the stability of the black hole increases, such that at $P\geq P_{c1}$, the black hole becomes completely stable.
In Figure \eqref{fig8}, we can observe a phase structure for this black hole for different values of $\alpha$. As $\alpha$ increases, the pressure and temperature values at critical points also increase.
\subsection{Case III}
We use equations \eqref{12} and \eqref{15} to obtain the temperature and eliminate the variable $P$, which results in,
\begin{equation}\label{28}
\begin{split}
T =\frac{4 \sqrt[3]{2} c_0^2 c_2 m^2 S^{2/3}+3\ 2^{2/3} c_0^3 c_3 m^2 \sqrt[3]{S}+2 c_0 c_1 m^2 S-4 e^2}{8 \pi  S}.
\end{split}
\end{equation}
Then, we obtained the thermodynamic function by inserting \eqref{28} into \eqref{16},
\begin{equation}\label{29}
\begin{split}
\Phi =\frac{4 \sqrt[3]{2} c_0^2 c_2 m^2 S^{2/3}+3\ 2^{2/3} c_0^3 c_3 m^2 \sqrt[3]{S}+2 c_0 c_1 m^2 S-4 e^2}{8 \pi  S \sin (\theta )}.
\end{split}
\end{equation}
Also, the vector field of the five-dimensional Yang-Mills black holes in massive gravity, according to relations \eqref{17} and \eqref{29}, is as follows,
\begin{equation}\label{30}
\begin{split}
&\phi^{S}=-\frac{\csc (\theta ) \left(2 \sqrt[3]{2} c_0^2 c_2 m^2 S^{2/3}+3\ 2^{2/3} c_0^3 c_3 m^2 \sqrt[3]{S}-6 e^2\right)}{12 \pi  S^2}\\
&\phi^{\theta}=-\frac{\cot (\theta ) \csc (\theta ) \left(4 \sqrt[3]{2} c_0^2 c_2 m^2 S^{2/3}+3\ 2^{2/3} c_0^3 c_3 m^2 \sqrt[3]{S}+2 c_0 c_1 m^2 S-4 e^2\right)}{8 \pi  S}.
\end{split}
\end{equation}
Therefore, we can calculate the critical points with respect to equation \eqref{30},
\begin{equation}\label{31}
\begin{split}
&S_1=-\frac{3 \left(9 c_3 c_0^2 m \left(2 c_2 e^2+c_3^2 c_0^4 m^2\right)+\sqrt{24 c_2 e^2+9 c_3^2 c_0^4 m^2} \left(2 c_2 e^2+3 c_3^2 c_0^4 m^2\right)\right)}{8 c_0^3 c_2^3 m^3}\\
&S_2=-\frac{3 \left(9 c_0^2 c_3 m \left(2 c_2 e^2+c_3^2 c_0^4 m^2\right)-\left(2 c_2 e^2+3 c_3^2 c_0^4 m^2\right) \sqrt{24 c_2 e^2+9 c_3^2 c_0^4 m^2}\right)}{8 c_0^3 c_2^3 m^3}.
\end{split}
\end{equation}
According to the above relationship, we can check the number of critical points in two cases. When $-\frac{3 c_0^4 c_3^2 m^2}{8 e^2}< c_2<0$ and $c_3>0$, we have two critical points, otherwise we will have only one or nothing critical point.
\begin{figure}[h!]
 \begin{center}
 \subfigure[]{
 \includegraphics[height=5cm,width=5cm]{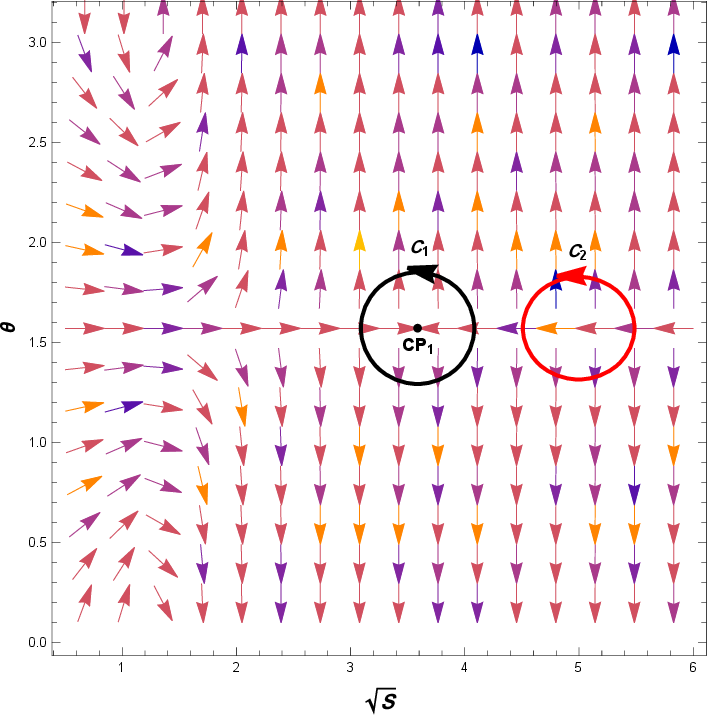}
 \label{fig9a}}
 \subfigure[]{
 \includegraphics[height=5cm,width=5cm]{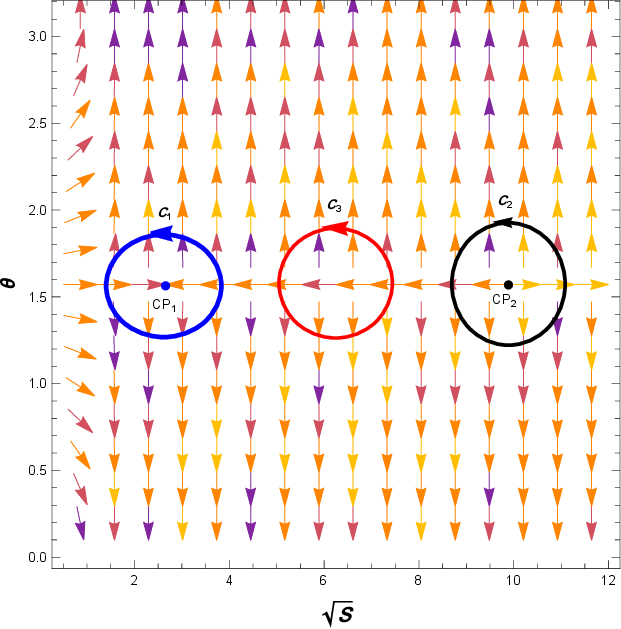}
 \label{fig9b}}
 \caption{\small{The  arrows represent the vector field n on  the $\sqrt{S}-\theta$ plane for the five-dimensional Yang–Mills black holes in massive gravity. So that, (a) $e=2$ and $m=c_0=c_1=c_2=c_3=1$, (b) $c_2=-0.3$ and $m=e=c_0=c_1=c_3=1$.}}
 \label{fig9}
 \end{center}
 \end{figure}
Also, the behavior of the deviation angle $\Omega(\vartheta)$, for the given contours in Figure \eqref{fig9}, is shown in Figure \eqref{fig10}.
 \begin{figure}[h!]
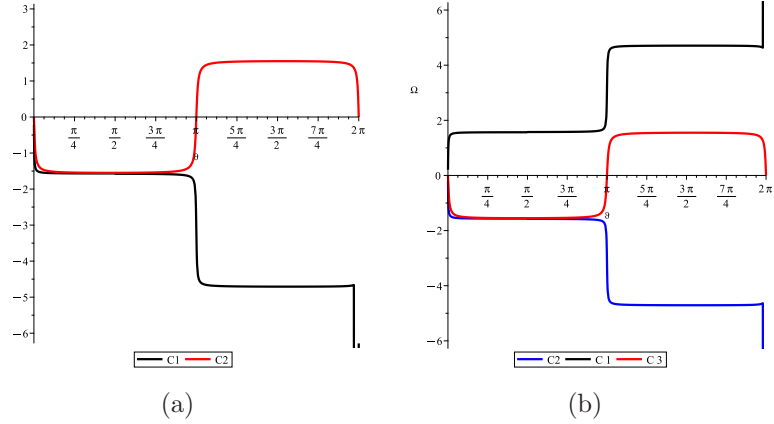

 \begin{center}
 \subfigure[]{
 \includegraphics[height=5cm,width=5cm]{fig4.eps}
 \label{fig10a}}
 \subfigure[]{
 \includegraphics[height=5cm,width=5cm]{fig5.eps}
 \label{fig10b}}
 \caption{\small{ $\Omega$ vs $\vartheta$ for contours $C$.}}
 \label{fig10}
 \end{center}
 \end{figure}
 As shown in Figure \eqref{fig9}, the values of $c_i$ (where $i=1,2,3$) determine the topological charges of the black hole. Therefore, massive gravity can affect these charges, as seen in Figure \eqref{fig9a} where we have a conventional topological charge and $Q_{total}=Q_{CP1}=-1$. In Figure \eqref{fig9b}, where the values of $c_i$ have changed, we have two conventional and novel topological charges, and $Q_{total}=Q_{CP1}+Q_{CP}=0$.
 \begin{figure}[h!]
 \begin{center}
 \subfigure[]{
 \includegraphics[height=5cm,width=5cm]{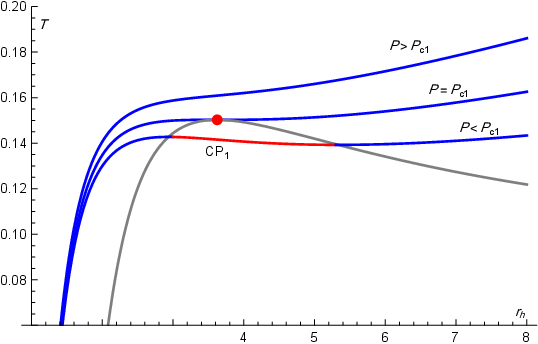}
 \label{fig11a}}
 \subfigure[]{
 \includegraphics[height=5cm,width=5cm]{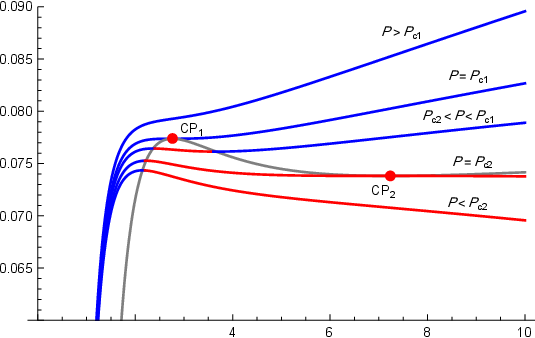}
 \label{fig11b}}
 \caption{\small{Isobaric curves (blue=stable, red=unstable) and spinodal curve (gray line)
for five-dimensional Yang–Mills black holes in massive gravity. We have set (a) $e=2$ and $m=c_0=c_1=c_2=c_3=1$, (b) $c_2=-0.3$ and $m=e=c_0=c_1=c_3=1$.}}
 \label{fig11}
 \end{center}
 \end{figure}
In Figure \eqref{fig11}, we can observe that the black hole is stable on both sides of the conventional critical point $CP_1$. However, in Figure \eqref{fig11b}, for the novel critical point $CP_2$, the black hole is unstable on both sides of this point. Therefore, we can conclude that phases (stability or instability) appear for critical points that have a topological charge of $+1$ ($Q_t=+1$), while phases disappear for critical points that have a topological charge of $-1$ ($Q_t=-1$).
\begin{figure}[h!]
 \begin{center}
 \subfigure[]{
 \includegraphics[height=5cm,width=5cm]{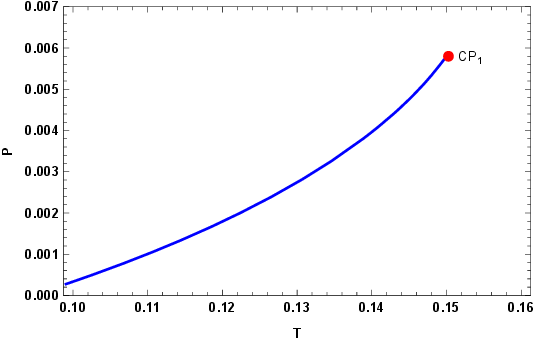}
 \label{fig12a}}
 \subfigure[]{
 \includegraphics[height=5cm,width=5cm]{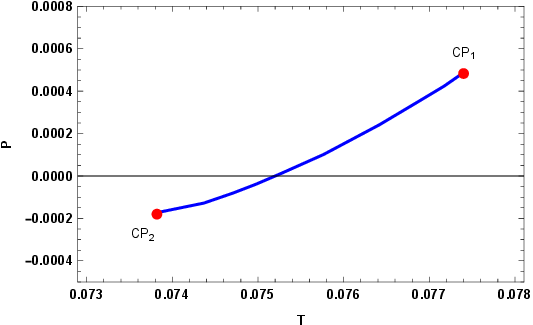}
 \label{fig12b}}
 \caption{\small{Phase diagram showing  for the five-dimensional Yang–Mills black holes in massive gravity. The first-order phase
transitions occur near the conventional critical point for (a) $e=2$ and $m=c_0=c_1=c_2=c_3=1$, (b) $c_2=-0.3$ and $m=e=c_0=c_1=c_3=1$ .}}
 \label{fig12}
 \end{center}
 \end{figure}
We have found that the five-dimensional Yang-Mills black hole in massive gravity belongs to two different topology classes. However, as shown in Figure \eqref{fig12}, this black hole exhibits the same phase structure behavior in both topological classes. In contrast, it has been suggested in \cite{23} that a black hole with a different topological class would have a different phase structure. An important finding is that massive gravity can alter the number of critical points of a Yang-Mills black hole, resulting in a different topology class. Nevertheless, it does not affect the phase structure of the Yang-Mills black hole. In Table 1, we summarize the topological charges associated with the critical points of the black hole under consideration. Additionally, the critical values can be found in Table 2.
\begin{center}
\begin{table}
  \centering
 \begin{tabular}{|p{4cm}|p{3.5cm}|p{3.5cm}|p{3cm}|}
   \hline
   \centering{Case} & \centering{Conditions} &  \centering{Topological Charge} & Total Topological Charge\\
 \hline
  \multirow{3}{5cm}{Euler-Heisenberg-AdS \newline black hole} &\centering{$0\leq a\leq \frac{32a^2}{7}$}  & $Q_{t|CP_1}=+1$,\newline $Q_{t|CP_2}=-1$ & $0$  \\
  & &  & \\
  & \centering{$a<0$} & $Q_{t|CP_1}=-1$ & $-1$ \\
   \hline
   Nonlinear \newline electrodynamic \newline black hole & \centering{$\alpha \geq 0$} & $Q_{t|CP_1}=-1$ & $-1$  \\[2mm]
   \hline
    \multirow{2}{5cm}{Five-dimensional\newline Yang–Mills black holes \newline in massive gravity} & $c_2\geq 0$,\newline $m,e,c_0,c_1,c_3>0$ &  $Q_{t|CP_1}=-1$  & $-1$ \\
      & &  & \\
    &$-\frac{3 c_0^4 c_3^2 m^2}{8 e^2}< c_2<0$,\newline  $m,e,c_0,c_1,c_3>0$ & $Q_{t|CP_1}=-1$ ,\newline $Q_{t|CP_2}=+1$ & $0$  \\
   \hline
 \end{tabular}
\caption{Summary of the results.}
\end{table}
 \end{center}

\begin{center}
\begin{table}
  \centering
 \begin{tabular}{|p{4cm}|p{3.5cm}|p{2.5cm}|p{1cm}|p{1cm}|p{1.5cm}|}
   \hline
   \centering{Case} & \centering{Conditions} &  \centering{Critical Point (CP)} & $r_c$& $T_c$&$P_c$\\
 \hline
  \multirow{3}{5cm}{Euler-Heisenberg-AdS \newline black hole} &$a=3$ & $CP_1$\newline $CP_2$ &1.5565 \newline 2.2876&0.0394 \newline 0.0444 & 0.0021 \newline 0.0035\\
  & &  & & &\\
  & $a=-1$ & $CP_1$ & 2.4865 &0.0430 &0.0032\\
   \hline
   \multirow{3}{5cm}{Nonlinear \newline electrodynamic \newline black hole} &$\alpha=0.1$ & $CP_1$ &2.2361 &0.0569  & 0.0087 \\
  & &  & & &\\
  & $\alpha=2$ & $CP_1$ & 1.0954 &0.4842 &0.1624\\
   \hline
    \multirow{2}{5cm}{Five-dimensional\newline Yang–Mills black holes \newline in massive gravity} & $c_2=1, e=2$,\newline $m=c_0=c_1=c_3=1$ &  $CP_1$  & 3.6235 & 0.1502&0.0058\\
      & &  & & &\\
    &$c_2=-0.3, e=1$,\newline  $m=c_0=c_1=c_3=1$ & $CP_1$ \newline $CP_2$ & 2.7639 \newline 7.2361 & 0.0774 \newline 0.0738 & 0.00048 \newline -0.00018\\
   \hline
 \end{tabular}
\caption{Critical values.}
\end{table}
 \end{center}

\section{Concluding remarks}
In this paper, we investigated a topological charge that is associated with critical points of Euler-Heisenberg, nonlinear, and Yang-Mills black holes in massive gravity.
We found that the number of critical points in the Euler-Heisenberg black hole is dependent on variable $a$. This leads to the classification of the black hole into two distinct topological classes, each with its own unique phase structure. The $Q_{total}=-1$  exhibits a small/large black hole phase structure, while the $Q_{total}=0$ displays a reentrant phase structure.
Also, if $a\rightarrow 0$, we get the same results as the charged AdS black hole \cite{5}. On the other hand, for the nonlinear electrodynamics black hole, we have only one conventional critical point, which has a small/large black hole phase structure and similar behavior to the charged AdS black hole. Additionally, we found that the massive gravity coefficient, $c_2$, can impact the number of critical points of the Yang-Mills black hole in massive gravity, resulting in the black hole being placed in two distinct topological classes. However, the Yang-Mills term does not affect the topological class. We also found that this black hole only has a small/large phase structure for different total topological charges. Here, the Euler-Heisenberg black hole and the Yang-Mills black hole in massive gravity have different topological classes, such as Born-Infeld AdS black holes in 4D novel Einstein-Gauss-Bonnet gravity \cite{22} and uncharged Lovelock AdS black holes \cite{23}. While the Yang-Mills black hole in heavy gravity is different from these black holes in terms of phase structure because it has a phase structure despite the different topology class. The obtained result is in conflict with the proposition \cite{23}, which states that black holes with different topological charges can have different phase structures. On the other hand, the novel and conventional critical points for the considered black holes correspond to the new classification \cite{21}. The way they categorize critical points is as follows: the novel critical point is where new phases (stable or unstable) appear as pressure increases, while the conventional point is where the phases disappear, as shown in Figs. \eqref{fig3}, \eqref{fig7} and \eqref{fig11}. We also found that, besides gauge corrections \cite{22}, massive gravity can alter the topological class. In addition, our findings align with the critical point parity conjecture presented in \cite{21}, which states that the total topological charge is odd (even) for an odd (even) number of critical points.
Finally, based on our results and \cite{21,22,23}  we conclude that the proposal presented in \cite{5}, which states the conventional critical points of the first phase transition, is a \emph{sufficient condition for black holes that have at most two critical points. However, if a black hole has more than two critical points, this proposition is only a necessary condition.}
\section{Acknowledgments}
This work is dedicated to the memory of Zohreh Sadeghi of the University of Washington, Tacoma, USA, (daughter of Prof.  J. Sadeghi). (For her diligent correction of the English texts of our papers).

\end{document}